\documentclass{aa}  
\usepackage{graphicx}
\usepackage{txfonts}

\newcommand\BATSE{{\it BATSE}}

\newcommand\hete{{\it HETE-II}}
\newcommand\swift{{\it Swift}}

\newcommand\kms{\ifmmode {\rm km\ s}^{-1} \else km s$^{-1}$\fi}
\newcommand\Hubble{\ifmmode {\rm km\ s}^{-1}\ {\rm Mpc}^{-1}
    \else km s$^{-1}$ Mpc$^{-1}$\fi}
\newcommand\ctssec{\ifmmode {\rm cts\ s}^{-1} \else cts s$^{-1}$\fi}
\newcommand\ergsec{\ifmmode {\rm ergs\ s}^{-1} \else
    ergs s$^{-1}$\fi}
\newcommand\eflux{\ifmmode {\rm ergs\ s}^{-1}\;{\rm cm}^{-2} \else
    ergs s$^{-1}$ cm$^{-2}$\fi}
\newcommand\phflux{\ifmmode {\rm photons\ s}^{-1}\;{\rm cm}^{-2}
    \else   photons s$^{-1}$ cm$^{-2}$\fi}
\newcommand\efluxA{\ifmmode {\rm ergs\ s}^{-1}\;{\rm cm}^{-2}\;{\rm
    \AA}^{-1} \else ergs s$^{-1}$ cm$^{-2}$ \AA$^{-1}$\fi}
\newcommand\efluxHz{\ifmmode {\rm ergs\ s}^{-1}\;{\rm cm}^{-2}\;{\rm
    Hz}^{-1} \else ergs s$^{-1}$ cm$^{-2}$ Hz$^{-1}$\fi}
\newcommand\cc{\ifmmode {\rm cm}^{-3} \else cm$^{-3}$\fi}
\newcommand\FWHM{\ifmmode {\rm FWHM} \else ${\rm FWHM}$\fi}
\newcommand\Msun{\ifmmode M_{\odot} \else $M_{\odot}$\fi}
\newcommand\Lsun{\ifmmode L_{\odot} \else $L_{\odot}$\fi}

\newcommand\Hbeta{\ifmmode {\rm H}\beta \else H$\beta$\fi}
\newcommand\Kalpha{\ifmmode {\rm K}\alpha \else K$\alpha$\fi}
\newcommand\NH{\ifmmode N_{\rm H} \else N$_{\rm H}$\fi}

\begin{document}

   \title{GRB\,090426: the farthest short gamma-ray burst?\thanks{The results reported in this paper are based on 
             observations carried out at Telescopio Nazionale Galileo and at Large Binocular Telescope}}

\author{
L. A. Antonelli \inst{1},
P. D'Avanzo\inst{2,3},
R. Perna \inst{4},
L. Amati\inst{6},
S. Covino\inst{2},
S. Cutini\inst{5},
V. D'Elia \inst{1,5},
S. Gallozzi \inst{1},
A. Grazian\inst{1},
E. Palazzi \inst{6},
S. Piranomonte\inst{1},
A. Rossi\inst{7},
S. Spiro\inst{1,8},
L. Stella\inst{1},
V. Testa\inst{1},
G. Chincarini\inst{3},
A. Di Paola\inst{1},
F. Fiore\inst{1},\\
D. Fugazza\inst{2},
E. Giallongo\inst{1},
E. Maiorano\inst{6},
N. Masetti\inst{6},
F. Pedichini\inst{1},
R. Salvaterra\inst{2},
G. Tagliaferri\inst{2},
and
S. Vergani\inst{9}
             }

   \offprints{L. A. Antonelli-- e-mail: a.antonelli@oa-roma.inaf.it }

   \institute{
INAF-Astronomical Observatory of Rome, via Frascati, 33, I-00040, Monteporzio (Rome), Italy
\and
INAF-Astronomical Observatory of Brera, via E. Bianchi, 46, Merate (LC), I-23807, Italy
\and
Universit\`a degli Studi di Milano "Bicocca", Piazza delle Scienze, 3, I-20126, Milano, Italy
\and
JILA, Campus Box 440, University of Colorado, Boulder, CO 80309-0440, USA
\and
ASI Science Data Center, Via Galileo Galilei, I-00044 Frascati (Roma), Italy
\and
INAF Ð Istituto di Astrofisica Spaziale e Fisica Cosmica, Via Gobetti 101, I-40126 Bologna, Italy
\and
Th\"uringer Landessternwarte Tautenburg, Sternwarte 5, 07778 Tautenburg, Germany
\and
Universit\'a degli Studi di Roma "Tor Vergata" - Dipartimento di Fisica - Via della Ricerca Scientifica, 1,  I-00133, Roma, Italy
\and
Laboratoire APC, Universit\`e Paris 7, 10 rue Alice Domon et Leonie Duquet, 75205 Paris Cedex 13, France
 }

 \date{Received on August 4$^{th}$, 2009 ; accepted on October 28$^{th}$, 2009}

 \abstract 
  {} 
  {With an observed and rest-frame duration of $< 2$s and $< 0.5$s, respectively, GRB\,090426 
  could be classified as a short GRB. The prompt detection, both from space and ground-based 
  telescopes, of a bright optical counterpart to this GRB offered a unique opportunity to complete a 
  detailed study.}  
  {Based on an extensive ground-based observational campaign, we obtained the spectrum of the optical afterglow of GRB\,090426, 
   measuring its redshift and obtaining information about the medium in which the event took place. We  completed follow-up observation of the afterglow optical 
   light curve down to the brightness level of the host galaxy that we firmly identified and studied. We also retrieved and analyzed all the available 
   high-energy data of this event, and compared the results with our findings in the optical. This represents one of the most 
   detailed studies of a short-duration event presented so far.
}  
  {The time properties qualify GRB\,090426 as a $short$ burst. In this case, its redshift of $z=2.61$ would be the highest yet found for a GRB of
  this class. On the other hand, the spectral and energy properties are more similar to those of $long$ bursts.
  LBT late-time deep imaging identifies a star-forming galaxy  at a redshift consistent with that of the GRB. 
  The afterglow lies within the light of its host and shows evidence of local absorption.}
  {}


  \titlerunning{The short burst GRB\,090426}
\authorrunning{L.A. Antonelli et al.}

   \maketitle
%

\section{Introduction.}

Gamma-ray bursts (GRBs) are intense gamma-ray flashes with typical duration ranging
from milliseconds to thousands seconds. Two classes of GRBs have been
identified: short-duration (less than 2 s) and hard-spectrum GRBs (SHBs), and long-duration 
(more then 2 s) and soft-spectrum GRBs (LSBs) (e.g.  \cite{Kouveliotou93}). For many years, 
the lack of observational data at both X-ray and optical wavelengths hampered the study of short 
GRBs. In 2005, an important breakthrough was provided by
the first detections of SHB afterglows triggered by $\swift$ and $\hete$. 
The  discovery of optical afterglows enabled the measurement of SHB redshifts and  
identification of their host galaxies (e.g., \cite{Berger05,Bloom06,DAvanzo09}). This established that, 
like LSBs, SHBs are also high-redshift relativistically expanding sources. In the past four years, it has been possible to derive a redshift for at least 10 SHBs. However, in all of those cases, the redshift was 
not directly measured, but inferred from the redshift of the putative host galaxy. In addition, the redshift 
distribution of SHBs appears to differ from that of LSBs, supporting the conjecture that they have a 
different origin. 
Unlike LSBs, SHBs are probably not produced by the collapse of massive stars (e.g., \cite{Nakar07} 
for a review). A popular model for SHBs involves the coalescence of two compact objects bound in a binary system (of double set of neutron stars, or a neutron star (NS) and a black hole (BH)). These systems have survived two 
supernova explosions, the resulting kick velocity having taken the systems significantly far away from their region 
of origin before the two compact objects merge. The delay between binary formation and merging depends strongly 
on the initial system separation, and theoretical works (\cite{Bel02,Bel06}) showed that a substantial 
fraction of the merging events should take place outside, or in the outskirts, of galaxies. 
In addition, double-neutron-star systems may be formed by exchange interactions in core-collapse globular 
clusters (\cite{Grindlay06}). If most short GRBs originate in merging binaries, the present sample of short 
GRBs of known redshift indicates that at least $\sim10\%$ of the merging binaries are formed in this way 
(\cite{Guetta09}). Firm spectroscopic measurements of SHBs, as well as the properties of their host galaxies, 
are of crucial importance in determining their origin. 

\section{GRB\,090426.}

GRB\,090426 was discovered by the $\swift$ coded-mask Burst Alert Telescope (BAT) 
on April $26^{th}$ 2009,  at $T_{burst}$=12:48:47 UT (\cite{Cummings2009}). Following the BAT trigger, the $\swift$ 
satellite automatically slewed to observe the GRB region with the on\,board narrow field instruments. 
The X-ray Telescope (XRT) began observing the field 84.6 seconds after the BAT trigger detecting an 
uncatalogued X-ray source within the BAT error circle (\cite{Cummings2009}). A candidate afterglow was
also detected by the UV-Optical Telescope (UVOT) on\,board $\swift$ in the {\it white, B, U} filters and 
marginally in the {\it V} filter at coordinates  $RA({\it J2000.0})= 12^{h}36^{m}18.07^{s}$, $Dec({\it J2000.0})= 32\degr59\arcmin09\farcs6$ but it was not detected in the UVOT UV filters (\cite{Oates09}). The optical afterglow observed by UVOT was also confirmed by several observations from other ground-based telescopes (e.g., \cite{Xin09,Yoshida09a,Yoshida09b,Olivares09,Mao09}). A redshift of 2.609 was measured first by Keck/LRSI (\cite{Levesque2009a,Levesque2009b}) and confirmed by other groups with the VLT (\cite{Thoene09}) and the TNG (this work).


\section{X-ray observations.}

\begin{figure}[ht!]
         \centering
         \includegraphics[width=9cm, angle=0]{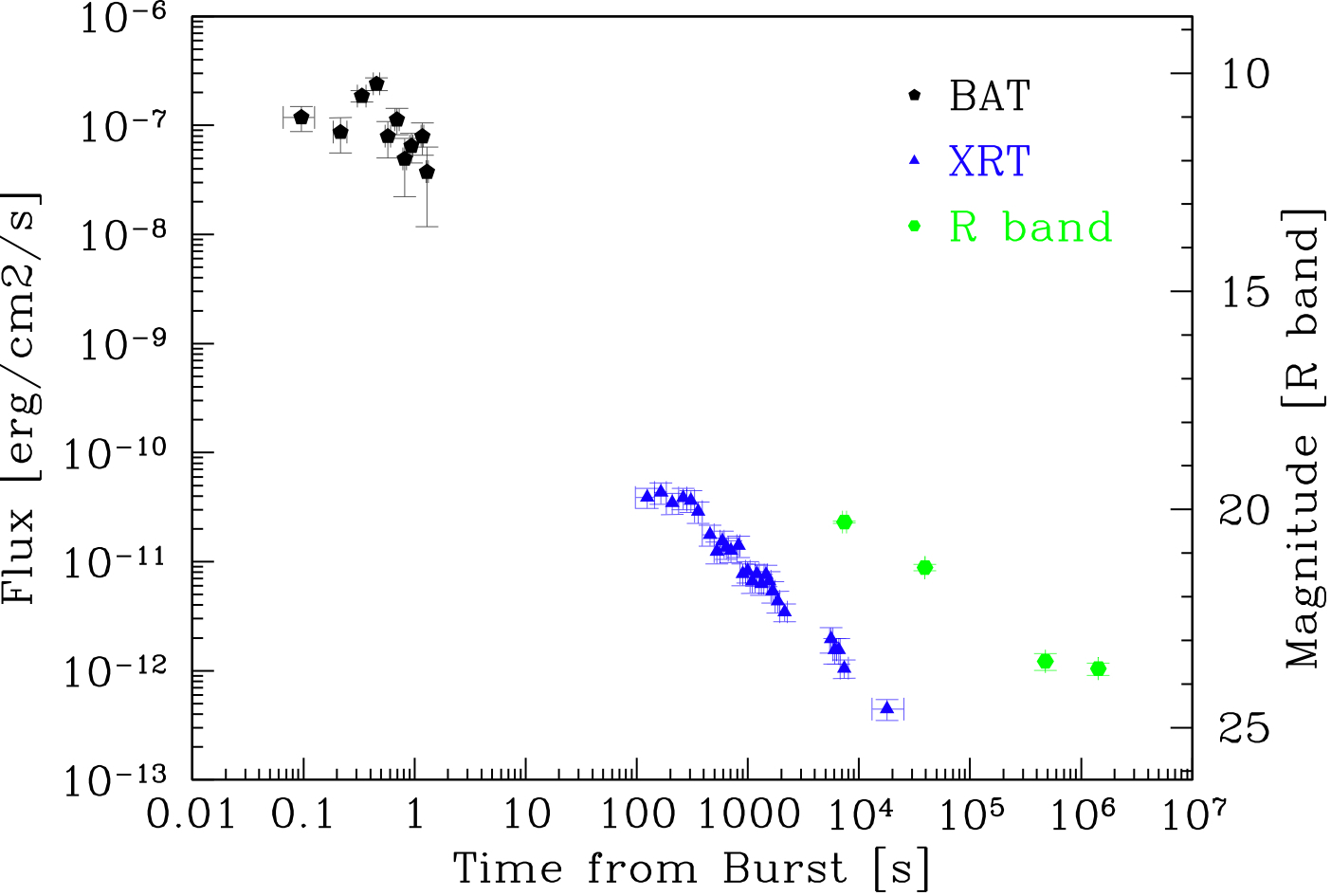}
                \caption{The X-ray light curve of GRB\,090426 extrapolated to the 0.3--10~keV band (filled pentagons) and
                the 0.3--10 keV band light curve of its X-ray afterglow (filled triangles).
                 Both X-ray light curves are background-subtracted and the time refers to the $\swift$ trigger.
                The $R$ band light curve (filled circles) from \cite{Mao09} (first point) and $TNG$. 
                }
                \label{xrtlcv}
        \end{figure}

The BAT event data were analyzed with the BAT analysis software included in the 
HEASOFT distribution (v.6.6.2) following standard procedures (\cite{krimm2004}).
All the errors reported are quoted at the 90\% confidence level. 
The burst has a duration of $T_{90}=1.25\pm0.25$ s in the $15-150$ keV band. 
The mask-weighted light curve shows a few overlapping peaks starting at about 
$T_{burst}-0.1$ s,  peaking at $T_{burst}+0.5$ s, and ending at $\sim T_{burst}+1.5$ s  
(Fig.\ref{xrtlcv}). 

The spectral lag of this GRB is consistent with zero with a relatively large uncertainty 
(\cite{Ukwatta2009}) and is consistent with the typical (negligible) temporal lag value measured 
for short GRBs (lags of few seconds are typical of long GRBs) (e.g., \cite{gehrels2006,norris00}). 
The BAT spectrum in the 15-150 keV energy band can be fit satisfactorily ($\chi^2=13.1/14$) by a simple 
power-law, yielding a photon index $\Gamma = 1.87_{-0.22}^{+0.23}$ and a fluence of $(1.7\pm0.2) \times10^{-7}$ erg cm$^{-2}$. 
Another fit with a Band function with photon indexes $\alpha$=-1, $\beta$ = -2.3 and the peak 
energy $E_p$ and the normalization as the only free parameters provides $E_p$=$49_{-18}^{+25}$ keV ($\chi^2=11.2/14$). 
By using this model and assuming the measured redshift of 2.61, we derive $E_ {\gamma,iso}= (5\pm1)\times10^{51}$ erg 
(in the 1-10000 keV cosmological rest-frame energy band) and $E_ {p,i} = 177_{-65}^{+90}$ keV\footnote{Assuming a flat 
Friedman-Robertson-Walker cosmology with $H_0 = 71$ km s$^ {-1}$ Mpc$^{-1}$, $\Omega_{\rm m} = 0.27$ and 
$\Omega_\Lambda = 0.73$}.  We note that this estimate of $E_p$ is fully consistent with the expected value of $\sim$50 keV 
based on the photon index -- peak energy correlation found for the SHB spectra detected by BAT (\cite{Sakamoto09}).  
 We also allow the $\alpha$ and $\beta$ parameters of the Band function to vary, obtaining 
 a lower limit to $E_p$ of 35 keV (90\% c.l.) and, in the cosmological rest frame of the source, 
 $E_{p,i} > 126$ keV, $E_{\gamma,iso}= [2.5-9.2]\times10^{51}$ ergs.\\
 The XRT observations started on April 26, 2009 ($\sim$85 s after the burst), and ended on 
May 5, 2009, thus adding up a total net exposure time (all obtained in photon-counting, 
PC mode) of about 54 ksec spread over 9 days. 
The monitoring was divided in 7 observations. The XRT data were processed with the 
{\it xrtpipeline} task (v0.9.9), applying standard calibration, filtering, and screening criteria.
At the beginning of the XRT observation, the intensity of the afterglow was high enough to cause 
pile-up in the data. To account for this effect, we extracted the source events in an annulus with a 
30-pixel outer radius and a 2-pixel inner radius. Then the entire circular region (30-pixel radius) 
was used to obtain late-time data. To account for the background, data were also extracted within 
an annular region (inner and outer radii of 50 and 100 pixels) centred on the source.  
All the errors reported are quoted at the 90\% confidence level. 
The extracted light curve is fitted well by a broken power-law with indices $\alpha_1=-0.20^{+0.01}_{-0.02}$ 
and $\alpha_2=-1.04^{+0.06}_{-0.08}$, and a break time of $T_{burst}=259$s.
Spectra of the source and background were extracted from the same regions described above during the first orbit. 
The (0.3 -- 10) keV band spectra were rebinned with a minimum of 20 counts per energy bin to allow the use 
of the $\chi^2$ statistic, and fitted by adopting a simple model of a galactic absorber plus an absorbed (at z=2.61) 
power-law model. The galactic hydrogen column density was fixed at $1.51\times10^{20}$ cm$^{-2}$ (\cite{kalberla05}). 
We found a spectral photon index of $\Gamma=2.04^{+0.49}_{-0.30}$ and an intrinsic equivalent hydrogen column density 
at z=2.61 of $N_{H}=2.3^{+5.6}_{-1.9}\times10^{21}$ cm$^{-2}$. 

\section{Optical observations.}

We observed the field of GRB\,090426 with the Italian 3.6-m TNG telescope, located in La Palma (Canary Islands), using
the DOLORES (Device Optimized for the LOw RESolution) camera, and with the Large Binocular Telescope (LBT), located on Mt.
Graham (Arizona), using the Large Binocular Cameras (LBC, both the blue and the red channel).

\begin{figure}
\centering
\includegraphics[width=10cm,angle=0]{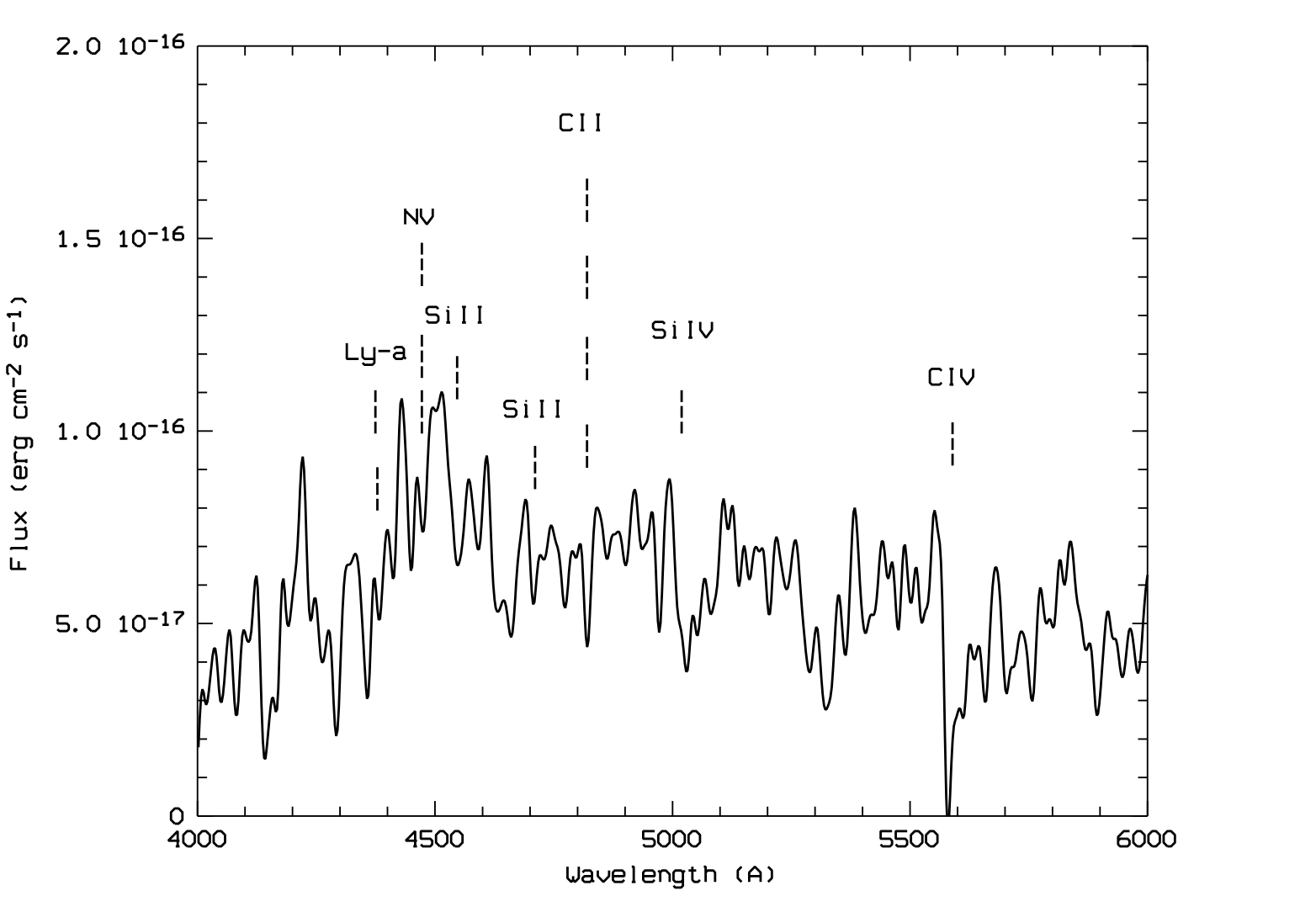}
   \caption{$DOLORES$spectrum of the optical afterglow of GRB\,090426 obtained at the
   TNG about 0.45 days after the burst. The redshift of $2.61\pm0.01$ is derived from the 
   observed lines.
    }
    \label{fig:optspe}
\end{figure}

The optical afterglow reported by Cummings et al. (2009) and Xin et al. (2009) was clearly visible in our $B$, $V$, $R$, and $I$ 
band images taken at TNG on April 27, 2009 (Table \ref{log}). 
Its coordinates $RA(J2000) = 12^{\rm h} 36^{\rm m}18\fs06$, $Dec(J2000) = +32\degr 59\arcmin 09\farcs3$ ($0\farcs3$ error) are fully consistent with the enhanced $\swift$-XRT and UVOT positions (\cite{Osborne09,Oates09}). 
During the same night, we also obtained medium-resolution spectra of the afterglow using the DOLORES camera with the grism 
LR-B and a $1\arcsec{}$ slit. The final reduced spectrum, smoothed using a 6 pixel width Gaussian filter, is shown in 
Fig. \ref{fig:optspe}; it has a signal-to-noise ratio of $5<S/N<10$ and covers the 3000--8000~\AA{}  wavelength range with a resolution of $R = 585$. 
From the detection of several absorption lines (Table\ref{lines}), we derived a redshift of $z = 2.61 \pm 0.01$. 
This is consistent with the measurements reported in  \cite{Levesque2009a,Thoene09}, and \cite{Levesque2009b}. 
From the optical spectrum, using {\sc FITLYMAN} within the MIDAS package, we also measured the column 
density of the Ly-$\alpha$ line at z=2.61, obtaining $N_{HI}=1.8^{+2.2}_{-0.8}\times10^{18}$ cm$^{-2}$. However, we note that
there is a second component (see Fig. \ref{fitspe}) that, if interpreted as Ly-$\alpha$ absorption at $z=2.59\pm0.01$,  has a 
column density of $N_{HI}=1.4^{+0.9}_{-0.5}\times10^{19}$ cm$^{-2}$ (uncertainties are given at 1$\sigma$). Both values are 
much lower than the typical column densities of GRB afterglows with Ly-$\alpha$ lines ($N_{HI}\sim10^{21}$ cm$^{-2}$) and are
consistent with the upper limit derived by Levesque et al., (2009b) from a Keck/LRIS spectrum. 
Owing to the poor S/N of the other lines, the metal column densities could not be derived. 
However, their large optical depths, together with the unusual low $N_H$ values, leads us to conclude that the host 
absorber should have a metallicity exceeding the solar one. This result is also confirmed by the higher $N_H$ derived 
from the X-ray data, whose value is obtained by taking into account the absorption edges of heavy elements based on
the assumption of solar abundance.
\begin{figure}[ht!]
\centering
\includegraphics[width=9cm]{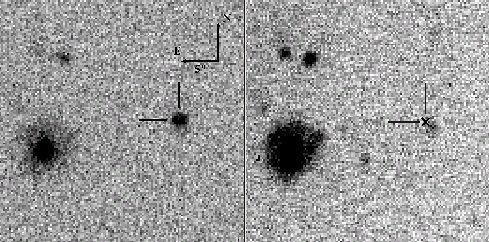}
   \caption{$R-$band images of the field of GRB\,090426 observed with the
   TNG about 0.4 (left) and 16.5 (right) days after the burst. In the right panel, a faint, extended 
   object is visible about $1.3''$ SW with respect to the optical afterglow position (marked by a cross).
     	 }
    \label{fig:fc}
\end{figure}

With the aim to search for the host galaxy (HG), we monitored the field at several epochs in different bands with both 
TNG and LBT (Table \ref{log}). The afterglow and the host galaxy were detected in the late TNG images but were not clearly resolved (Fig. \ref{fig:fc}). 
Deep LBT images were obtained in the four SDSS-like filters $g'$, $r'$, $i'$, and $z'$ between May 30, 2009 and June 1, 2009 
using the LBC prime-focus camera, in binocular mode as reported in Table \ref{log}. 
In the LBT images, we observed a clearly extended object with an irregular shape that could be the host galaxy (HG) of GRB\,090426. 
The emission maximum of this object occurs at $RA(J2000) = 12^{\rm h} 36^{\rm m} 17\fs98$, 
$Dec(J2000) = +32\degr 59\arcmin 08\farcs6$, i.e., $\sim1\farcs4$ S-W from the optical afterglow position (Fig.~\ref{fig:optlbc}). 
At the position of the GRB afterglow we observed an extended emitting knot (hereafter N-E knot) (Table \ref{log}) . 
Both objects were also studied in Levesque et al. (2009b). 
This extended emission appears to be connected with the HG suggesting that they are either part of the same object (e.g., a $HII$ region) or an interacting system. From our multiband $g',r',i',z'$ optical photometry, using the $z-phot$ program described in \cite{Fontana00}, we derived for the observed object a photometric redshift of $z=2.28\pm0.35$, which is consistent (within the errors) with the measured redshift of the afterglow. If we assume that this object is the HG of GRB\,090426, then its $B-$band absolute magnitude is $-23.6\pm0.2$ and the afterglow would be located at a projected distance of $\sim 10.6$ kpc from the emission maximum. 

\section{Results and discussion}

\begin{figure}[ht!]
         \centering
         \includegraphics[width=9cm, angle=0]{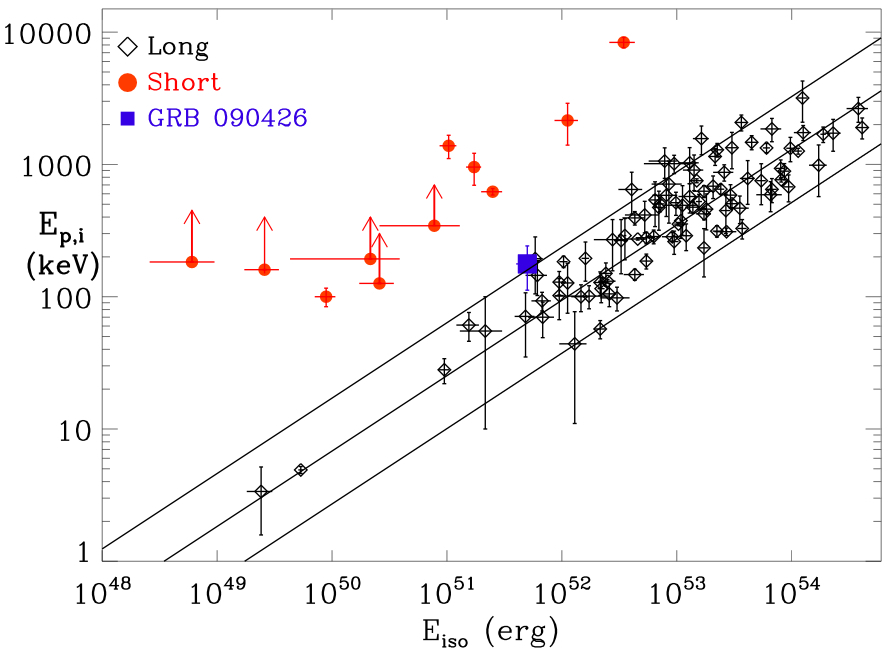}
                \caption{Location of GRB\,090426 (red filled square) in the $E_{p,i} - E_{iso}$ plane. The data points of long GRBs (black open diamonds) are from 
                \cite{Amati08} and \cite{Amati09}, the data points and limits of short GRBs (red filled dots) are from \cite{Amati08, Pirompo08,Amati09}. 
                The continous lines show the best fit power--law and the $\pm2\sigma$ confidence region of the correlation, as determined by \cite{Amati08}.}
                \label{amatirel}
        \end{figure}

 Most of the properties of SHB prompt emission are derived from \BATSE\,  
 bursts by using the sampling criterion $T_{90}<2$ s. This distinction suffices for the 
 purpose of statistical population analysis, but the 2\,s duration dividing-line gives rise 
 to a significant overlap between long and short GRBs. To classify a SHB, 
 additional information about the spectrum, released energy, and temporal structure are required. 
 GRB\,090426 had a $T_{90}=1.25\pm0.25$~s (corresponding to 
 $T_{90}^{rf}=0.35\pm0.07$~s in its rest-frame), which is well within the range typical of SHBs. 
 The isotropic energy released in the prompt phase, $E_{\gamma,iso}= (5\pm1)\times10^{51}$ erg, 
 remains consistent with the typical energy of SHBs, though close to the higher end of the range (e.g., \cite{Nakar07}).  
 The hardness ratio at high energies shows that short bursts are on average harder than the long GRBs (e.g., \cite{Kouveliotou93}). 
 The spectrum of GRB\,090426 is well within the photon index distribution of BAT spectra of \swift\ SHBs, even if it is one 
 of the softest observed (\cite{Sakamoto09}). Nevertheless, we note that the estimated values (or limits) of $E_{p,i}$ and $E_{iso}$ of GRB\,090426 lie at the border of the $2\sigma$ confidence region of the $E_{p,i}$ -- $E_{iso}$ correlation that holds for 
 LSBs (Fig. \ref{amatirel}) (\cite{Amati08}). This is quite peculiar because other SHBs of known redshift and measured 
 $E_{p,i}$ are outliers to this correlation at more than 3$\sigma$ c.l. (Fig. \ref{amatirel}) (e.g., \cite{Pirompo08}). 
 From the point of view of its spectral and energetic properties, GRB\,090426 straddles the SHB and LSB classes. 
 Based on the burst duration and peak energy of GRB 090426, Levesque et al (2009b) classify 
this GRB as a SHB. However,  when considering the large number of LSB detected so far, these authors also note that the 
combination of parameters leading to this classification might have occurred by chance from the distribution of parameters
 characterising LSB. Therefore, even though the 0.3\,s duration would place GRB\,090426 in the short burst category, its classification remains uncertain.
  The X-ray afterglow light curve shows the typical power-law decay of
 SHBs ($F\propto$ $t^{-\alpha}$) with a short initial plateau in the
 first $260$ s followed by a smooth decay ($\alpha_x=1.04^{+0.06}_{-0.08}$) 
 without further breaks until $5\times10^5$ s after the burst (Fig.~\ref{xrtlcv}).  
 The initial flattening is not common in the X-ray afterglow of SHBs but it
 was observed in at least another case:  GRB 061201 (\cite{Stratta07}). 
 The optical light curve of the afterglow shows a power--law decay 
 with a slope that is shallower than in X-ray ($\alpha_o=0.74\pm0.03$).  
Such a steeper decay in the X-ray band is roughly consistent with the prediction of the 
standard fireball model for an outflow propagating in a constant-density circumburst medium 
when the cooling break occurs between the optical and X-ray bands.
Later, at $T-T_{burst}\sim22$ days, the $R$-band optical light curve 
became flatter. This flattening was probably caused by the emerging flux of the 
HG, which is not resolved well in the TNG images but is clearly 
detected in the late-time LBT observations. 
The X-ray spectrum has a typical value of the spectral index but an intrinsic absorption of 
$N_{H}=2.3^{+5.6}_{-1.9}\times10^{21}$ cm$^{-2}$ in the rest-frame which is quite high for SHBs. 
This high intrinsic absorption derived from the spectrum of the X-ray afterglow implies a 
dense circumburst region as often observed for LSBs. This is also confirmed 
by the presence of highly ionized absorption lines such as NV and CIV and SiIV in the optical 
spectrum. To study the HG 
we performed a photometric fitting of the LBT data of both the HG and the N--E knot assuming $z=2.61$ 
and using different synthetic galaxy templates and extinction laws. A large number of models provide 
a good fit for the HG but the best fit ($\chi^{2}=3.55\,(3\, d.o.f.)$) is provided by an evolved galaxy model 
with a {\it Milky Way} extinction law ($E(B-V)=0.20^{+0.25}_{-0.05}$) and a specific star formation rate 
(SSFR: the ratio of SFR to the stellar mass of the galaxy) of $(79^{+24}_{-37})\times10^{-3}$ Gyr$^{-1}$.
We also found, with a larger uncertainty, that the N--E knot (associated with the GRB location) is 
characterized more accurately by a higher extinction ($E(B-V)=0.6_{-0.3}^{+0.15}$) and a higher SFR. This knot could be a 
star-forming region possibly related to the HG or a different object interacting with the HG (Fig. \ref{fig:optlbc}).

Regardless of its classification, GRB\,090426 represents an intriguing example of GRB progenitors. 
In the scenario involving the merging of a double compact object system, the position of the afterglow 
within the host galaxy, the intrinsic absorption measured in the X-ray spectrum, and the detection of highly ionized 
absorption lines (such as NV, CIV and SiIV) in the optical spectrum of the afterglow are strongly indicative of a
``primordial'' binary system that merged in a relatively short time ($10^7 - 10^8$ yr), when 
most systems remained in side their star-forming regions. The existence of these ``short-lived'' 
systems has been investigated by several authors (e.g. \cite{Bel01,Perna02,Bel06}), and the association of this progenitor 
with GRB\,090426 is also in close agreement with the relatively high redshift of this event (the highest measured for a 
short-duration GRB). On the other hand, the same properties listed above (low offset, intrisic absorption, high$-z$), 
together with the relatively high SFR measured for the host galaxy (Sect. 3.1) and the consistency with the 
$E_{p,i}$ -- $E_{iso}$ correlation, can also be interpreted as well by assuming a core-collapse supernova progenitor, 
as it has been observed for many long GRBs. But it is then not easy to reconcile the short rest-frame 
duration of the prompt event ($< 1$s) with the expected size of the progenitor,  which, otherwise, makes the association 
of GRB\,090426 with a massive star progenitor not straightforward (e.g.,  \cite{Woosley06}).

\begin{acknowledgements}
We acknowledge an anonymous referee for useful comments and the excellent 
support from TNG staff. LAA  thanks F. Calabr\'o and M. Lopergolo for having made 
his research possible. The LBT is an international collaboration among institutions 
in the USA, Italy, and Germany. LBT Corporation partners are the University
of Arizona, on behalf of the Arizona university system;
Istituto Nazionale di Astrofisica, Italy; LBT Beteiligungsgesellschaft,
Germany, representing the Max Planck Society, the Astrophysical
Institute Potsdam, and Heidelberg University; The Ohio
State University; and the Research Corporation, on behalf of the
University of Notre Dame, University of Minnesota, and University
of Virginia. 
\end{acknowledgements}

\Online

\begin{appendix} 
\begin{table*}
\caption{Observation log for GRB\,090426. Magnitudes are in the AB system and are not corrected for
Galactic extinction (E(B-V)= 0.02 mag \cite{Schlegel98}). Point spread function (PSF) and aperture
photometry were performed, on both the afterglow and the host galaxy respectively, by using the 
{\it Daophot II} within the ESO-MIDAS package, the task Phot/Apphot within the IRAF 
package and the SExtractor package (\cite{sex96}). In particular, an aperture radius of 2.2" was adopted 
to cover the entire extended object observed in the LBT images.}
The calibration  of TNG data was performed using Landolt standard stars.  The calibration of LBT was done 
against the SLOAN catalogue. Errors and upper limits are given at $1\sigma$ confidence level.
\centering
\begin{tabular}{cccccccc} \hline
Mean time          &  Exposure time             & Time since GRB &    Seeing &  Instrument       &  Magnitude               & Filter       &   Comments  \\
(UT)                     &  (s)                                   & (days)                 & (\arcsec)  &                             &                                    & Grism     &	        \\ 
\hline
2009 Apr   26.97190 &  $ 4 \times 120$	   &  0.43802	   & 1.4       & TNG/DOLORES	& $22.05 \pm 0.08$   & $B$         & AG detection   \\
2009 Apr   26.97679 &  $ 2 \times 120$	   &  0.44291	   & 1.4       & TNG/DOLORES	& $21.69 \pm 0.07$   & $V$         & AG detection   \\
2009 Apr   26.98818 &  $ 1 \times 180$	   &  0.45430	   & 1.4       & TNG/DOLORES	& $21.38 \pm 0.06$   & $R$        & AG detection   \\
2009 Apr   26.99583 &  $ 1 \times 180$	   &  0.46195	   & 1.4       & TNG/DOLORES	& $21.25 \pm 0.09$   & $I$          & AG detection   \\ 
\hline
2009 Apr   26.98457 &  $ 2 \times 1800$	   &  0.45069	   & 1.4       & TNG/DOLORES	& $-$		          & $LR-B$  & AG spectra     \\ 
\hline
2009 May  02.04516 &  $16 \times 120$	   &   5.51128	   & 1.4       & TNG/DOLORES	& $23.54 \pm 0.16$   & $R$        & HG detection   \\
2009 May  13.05282 &  $30 \times 180$	   &  16.51894	   & 1.0       & TNG/DOLORES	& $23.70 \pm 0.12$   & $R$        & HG detection   \\
2009 May  18.97929 &  $20 \times 180$	   &  22.44541	   & 1.0       & TNG/DOLORES	& $> 24.0$	          & $I$          & $3\sigma$ u.l. \\ 
\hline
2009 May  31.20833 &  $21\times 180$       &  35.67445      & 0.9        & LBT/LBC\_blue     & $24.11 \pm 0.10$  & $g'$        & HG + knot \\
2009 May  31.21875 &  $17\times 180$       &  35.68487      & 1.1        & LBT/LBC\_blue     & $23.83 \pm 0.13$  & $r'$         & HG + knot \\
2009 May  31.20833 &  $16\times 180$       &  35.67445      & 0.7        & LBT/LBC\_red       & $24.06 \pm 0.18$  & $i'$         & HG + knot  \\
2009 May  31.21875 &  $12\times 180$       &  35.68487      & 0.8        & LBT/LBC\_red       & $23.32 \pm 0.15$  & $z'$        & HG + knot  \\ 
\hline
2009 May  31.20833 &  $21\times 180$       &  35.67445      & 0.9        & LBT/LBC\_blue     & $24.70 \pm 0.10$  & $g'$        & HG  \\
2009 May  31.21875 &  $17\times 180$       &  35.68487      & 1.1        & LBT/LBC\_blue     & $23.86 \pm 0.07$  & $r'$         & HG  \\
2009 May  31.20833 &  $16\times 180$       &  35.67445      & 0.7        & LBT/LBC\_red       & $24.15 \pm 0.13$  & $i'$         & HG  \\
2009 May  31.21875 &  $12\times 180$       &  35.68487      & 0.8        & LBT/LBC\_red       & $23.73 \pm 0.13$  & $z'$        & HG  \\ 
\hline
2009 May  31.20833 &  $21\times 180$       &  35.67445      & 0.9        & LBT/LBC\_blue     & $25.46 \pm 0.20$  & $g'$        & N-E knot \\
2009 May  31.21875 &  $17\times 180$       &  35.68487      & 1.1        & LBT/LBC\_blue     & $24.72 \pm 0.15$  & $r'$         & N-E knot \\
2009 May  31.20833 &  $16\times 180$       &  35.67445      & 0.7        & LBT/LBC\_red       & $25.53 \pm 0.32$  & $i'$         & N-E knot \\
2009 May  31.21875 &  $12\times 180$       &  35.68487      & 0.8        & LBT/LBC\_red       & $24.87 \pm 0.29$  & $z'$        & N-E knot \\ 
\hline
\end{tabular}
\label{log}
\end{table*}

\begin{table*}
\caption{Rest-frame equivalent widths and 1-$\sigma$ errors of the absorption lines detected in the TNG spectrum. }
\centering
\begin{tabular}{lcc} \hline
Line &                     EW  &   $\Delta$EW \\
 & (\AA) &     (\AA)\\
\hline
Ly-$\alpha$ (z=2.61) &                   0.8  &  0.3      \\
Ly-$\alpha$ (z=2.58) &                   2.8  &  0.3      \\
NV$\, \lambda \,1238, 1242$ &     0.4  &  0.3      \\
SiII$\, \lambda \,1260  $           &    1.3  &  0.3      \\
SiII$\, \lambda \,1304 $            &    0.9  &  0.3      \\
CII$\, \lambda \,1334 $             &     1.3 &  0.3 \\
SiIV$\, \lambda \,1393, 1402$  & 7.0      &  0.3\\
CIV$\, \lambda \,1548, 1550$   & 7.5       &  0.3\\
\hline
\end{tabular}
\label{lines}
\end{table*}
\end{appendix}

\begin{appendix} 


\begin{figure*}
\centering
\includegraphics[width=10.cm, clip]{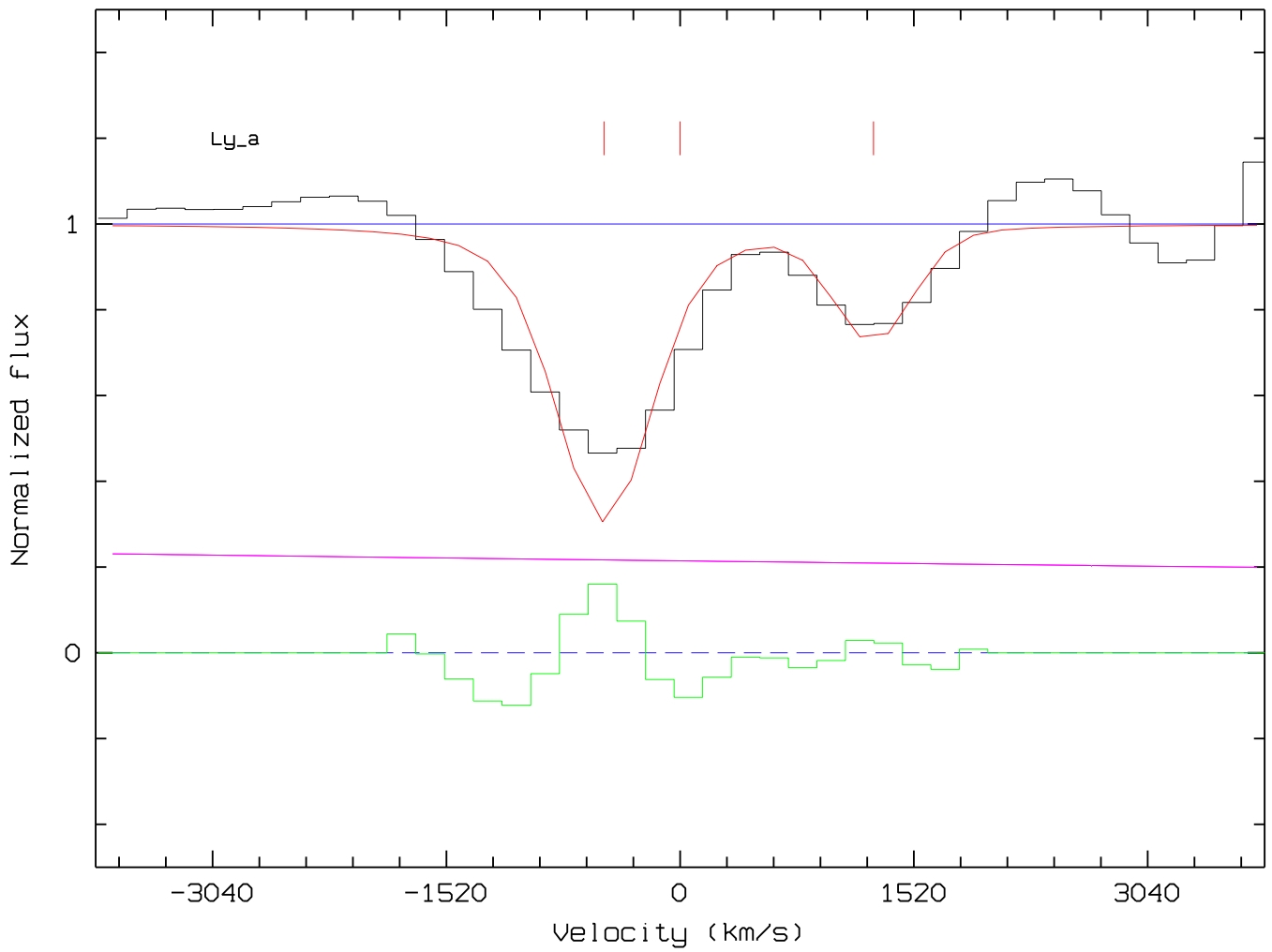}
   \caption{Voigt profile fitting of the Ly-$\alpha$ absorption feature in the GRB090426 spectrum.
                  Two different components are identified: the first is at z=2.61 and consistent with the other lines, the second 
                   is at a different redshift z=2.59. The central tick marks the arbitrary x-axis origin.}
    \label{fitspe}
\end{figure*}

\begin{figure*}
\centering
\includegraphics[width=16.4cm, clip]{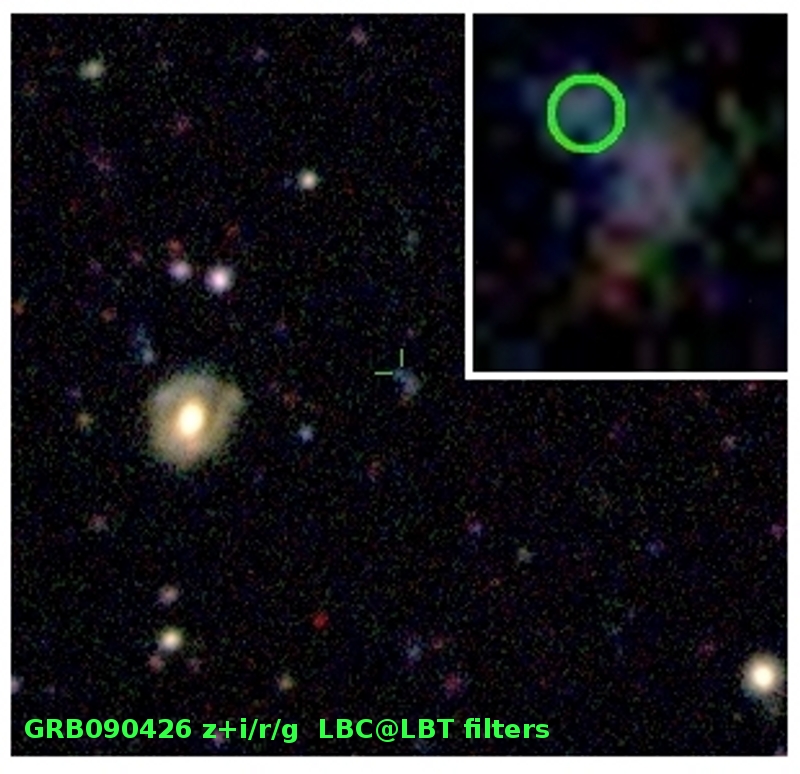}
   \caption{$Large\,Binocular\,Camera\,Blue \& Red\,Channels$ image of the host galaxy of
   GRB\,090426 obtained at LBT about 35 days after the burst. The host galaxy of the GRB
   appears irregular and the core is located $\sim1.4"$ S-W from the afterglow position (marked by the two lines).
   Coincident with the afterglow position (green $0\farcs4$ radius error circle), there is an emitting knot that should 
   be part of the host galaxy. The images were processed by using the LBC reduction pipeline at the LBC Survey 
   Center at the Astronomical Observatory of Rome. The images were calibrated with a subcatalog extracted from the 
   SDSS, by selecting the highest S/N unsaturated objects. LBC fit mosaic to any filter was aligned and resampled at the same 
   pixel grid, then convolved to the a same common seeing; filters were associated wiht RGB channels according to this sequence: 
   red channel as a composition of (z+i)/2 SLOAN filters, green as r SLOAN filter and blue as g SLOAN filter. 
   Resulting fitted images became the input of a rendering colour program (we used STIFF), to examine the input images and create     
   a histogram of pixel values, from which we derive statistics: low and high cuts were determined automatically to optimize the 
   dynamic gamma range using quantile at 95\%. The gamma factor is also interpolated for each pixel using a luminance factor of Y=(R+G+B)/3.
   }
    \label{fig:optlbc}
\end{figure*}

\begin{figure*}
\centering
\includegraphics[width=16.5cm, clip]{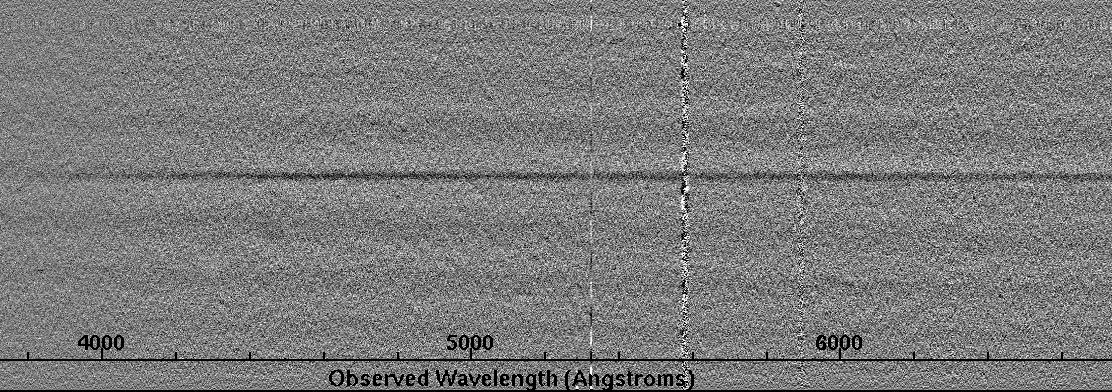}
   \caption{ Two-dimensional TNG+DOLORES spectrum of GRB\,090426.}
    \label{2dspe}
\end{figure*}


\end{appendix}


\begin{thebibliography}{}

\bibitem[Amati et al. 2008]{Amati08}
Amati, L., Guidorzi, C., Frontera, F., et al. 2008, MNRAS, 391, 577

\bibitem[Amati et al. 2009]{Amati09}
Amati, L., Frontera, F. and, Guidorzi, C., 2009, A\&A, in press.



\bibitem[Belczynski \& Kalogera 2001]{Bel01}
Belczynski, K. \& Kalogera, V., 2001, ApJ, 550, L183.

\bibitem[Belczynski et al. 2002]{Bel02}
Belczynski, K., Bulik, T., and Rudak, B. 2002, ApJ, 571, 394

\bibitem[Belczynski et al. 2006]{Bel06}
Belczynski, K., Perna, R., Bulik, T. et al., 2006, ApJ, 648, 1110

\bibitem[Berger et al. 2005]{Berger05}
Berger, E., Price, P. A., Cenko, S. B. et al. 2005, Nature 438, 988

\bibitem[Berger et al. 2007]{Berger07}
Berger, E., Fox, D., Price, P. A., et al., 2007, \apj\,, 664, 1000

\bibitem[Bertin \& Arnouts 1996]{sex96}
Bertin \& Arnouts, 1996, A\&AS, 117, 393)

\bibitem[Bloom et al. 2006]{Bloom06}
Bloom, J. S., Prochaska, J. X., Pooley, D. et al., 2006, ApJ, 638, 354

\bibitem[Cummings et al. 2009]{Cummings2009} 
Cummings, J.R., et al., 2009, GCN Circular, 9254

\bibitem[D'Avanzo et al. 2009]{DAvanzo09}
D'Avanzo, P.,  Malesani D., Covino, S., et al., 2009, A\&A, 498, 771


\bibitem[Fontana et al. (2000)]{Fontana00} 
Fontana, A., D'Odorico, S., Poli, F., et al.\ 2000, \aj, 120, 2206 

\bibitem[Gal-Yam et al. 2008]{Gal08} 
Gal-Yam, A., et al., 2008, \apj, 686, 408 

\bibitem[Gehrels et al. 2006]{gehrels2006} 
Gehrels, N. et al. 2006, Nature, 444, 1044



\bibitem[Guetta \& Stella, 2009]{Guetta09}
Guetta, D. \& Stella, L., 2009, A\&A, 498, 329

\bibitem[Grindlay et al. 2006]{Grindlay06}
Grindlay, J., Portegies Zwart, S., and McMillan, S., 2006, Nature Physics, 2, 116

\bibitem[Kalberla et al. 2005]{kalberla05} 
Kalberla, P.M.W., Burton, W.B., Hartmann, D., et al. 2005, \aap, 440, 775

\bibitem[Kouveliotou et al. 1993]{Kouveliotou93}
Kouveliotou, C., Meegan, C.~A., Fishman, G.~J., et al., 1993, ApJ,  413, L101.

\bibitem[Krimm et al. 2004]{krimm2004} 
Krimm, H.A. 2004, BAT Ground Analysis Software Manual

\bibitem[Levesque et al. (2009a)]{Levesque2009a} 
Levesque, E., et al., 2009, GCN Circular, 9264

\bibitem[Levesque et al. (2009b)]{Levesque2009b} 
Levesque, E.,  et al., 2009, MNRAS, submitted (arXiv:0907.1661)

\bibitem[Mao et al. 2009]{Mao09} 
Mao, J., Cha, G., Bai, J., 2009, GCN Circular, 9285



\bibitem[Nakar  2007]{Nakar07} 
Nakar, E.,  2007, Phys. Rev., 442, 166
 
\bibitem[Norris et al. 2000]{norris00}
Norris, J.~P., Marani, G.~F., and Bonnell, J.~T., 2000, ApJ, 534, 248


\bibitem[Oates et al. 2009]{Oates09}
Oates, S.R., et al., 2009, GCN Circulars, 9265

\bibitem[Olivares et al. 2009]{Olivares09} 
Olivares, F., et al 2009, GCN Circular, 9268

\bibitem[Osborne et al. 2009]{Osborne09}
Osborne, J.P., et al., 2009, GCN Circulars, 9259

\bibitem[Perna \& Belczynski 2002]{Perna02}
{Perna}, R. \& {Belczynski}, K. 2002, ApJ, 570, 252.

\bibitem[Piranomonte et al. 2008]{Pirompo08}
Piranomonte, S., D'Avanzo, P., Covino, S., et al., 2008, A\&A, 491, 183

\bibitem[Sakamoto et al. 2009]{Sakamoto09}
Sakamoto, T., Sato, G., Barbier, L. et al., 2009, ApJ, 693, 922




\bibitem[Schegel et al. 1998]{Schlegel98} 
Schlegel, D.J., Finkbeiner, D.P., \& Davies, M. 1998, \apj, 500, 525



\bibitem[Stratta et al., 2007]{Stratta07} 
Stratta, G., D'Avanzo, P., Piranomonte, S., et al., 2007, A\&A, 474, 827

\bibitem[Th\"one et al. (2009)]{Thoene09} 
Th\"one C., et al. 2009, GCN Circular, 9269

\bibitem[Ukwatta et al. 2009]{Ukwatta2009} 
Ukwatta, T.N., et al. 2009, GCN Circular, 9272

\bibitem[Xin et al. 2009]{Xin09} 
Xin, L.P., et al. 2009, GCN Circular, 9255

\bibitem[Yoshida et al. 2009a]{Yoshida09a} 
Yoshida, M., et al 2009a, GCN Circular, 9266

\bibitem[Yoshida et al. 2009b]{Yoshida09b} 
Yoshida, M., et al 2009b, GCN Circular, 9267

\bibitem[Woosley \& Bloom, 2006]{Woosley06}
Woosley, S.E. \& Bloom, J.S., 2006, \araa, 44, 507

\end{thebibliography}
\end{document}